\documentclass[journal]{ieeetran}
\usepackage{cite}
\usepackage{amsmath,amssymb,amsfonts}
\usepackage{algorithmic}
\usepackage{graphicx}
\usepackage{textcomp}
\usepackage{subcaption}
\usepackage[affil-it]{authblk}

\usepackage{hyperref}
\DeclareMathOperator{\Tr}{Tr}
\DeclareMathOperator\arctanh{arctanh}

\begin{document}
\renewcommand\Authfont{\normalfont}
\renewcommand\Affilfont{\itshape\small}
\title{Single-Qubit Fidelity Assessment of Quantum Annealing Hardware}
\author[1]{Jon Nelson}
\author[1, 2]{Marc Vuffray}
\author[1, 2]{Andrey Y. Lokhov}
\author[1]{Carleton Coffrin}
\affil[1]{Advanced Network Science Initiative, Los Alamos National Laboratory, Los Alamos, NM 87545 USA}
\affil[2]{Theoretical Division, Los Alamos National Laboratory, Los Alamos, NM 87545 USA}

\maketitle
\begin{abstract}
As a wide variety of quantum computing platforms become available, methods for assessing and comparing the performance of these devices are of increasing interest and importance.
Inspired by the success of single-qubit error rate computations for tracking the progress of gate-based quantum computers, this work proposes a Quantum Annealing Single-qubit Assessment (QASA) protocol for quantifying the performance of individual qubits in quantum annealing computers.
The proposed protocol scales to large quantum annealers with thousands of qubits and provides unique insights into the distribution of qubit properties within a particular hardware device.
The efficacy of the QASA protocol is demonstrated by analyzing the properties of a D-Wave 2000Q system, revealing unanticipated correlations in the qubit performance of that device.
A study repeating the QASA protocol at different annealing times highlights how the method can be utilized to understand the impact of annealing parameters on qubit performance.  
Overall, the proposed QASA protocol provides a useful tool for assessing the performance of current and emerging quantum annealing devices.
\end{abstract}

\section{Introduction}
\label{sec:introduction}

In the current era of Noisy Intermediate-Scale Quantum (NISQ) \cite{Preskill2018quantumcomputingin} devices, measuring and tracking changes in the fidelity of quantum hardware platforms is essential to understanding the limitations of these devices and quantifying progress as these platforms continue to improve.
Measuring the performance of gate-based quantum computers (QC) has been studied extensively through the topics of quantum characterization, verification, and validation (QCVV) \cite{Eisert2020}.
The scope of QCVV is broad and ranges from testing individual quantum operations (e.g., error rates of one- and two-qubit gates \cite{Wright2019}), verifying small circuits (e.g., Randomized Benchmarking \cite{PhysRevLett.106.180504,PhysRevA.77.012307}, Gate Set Tomography \cite{2009.07301}), to full system-level protocols (e.g., quantum volume estimation \cite{PhysRevA.100.032328}, random quantum circuits \cite{Boixo2018}).
Over the years these QCVV tools have become an invaluable foundation for benchmarking and measuring progress of quantum processors \cite{ibm_qv}, culminating with a quantum supremacy demonstration in 2019  \cite{Arute2019}.

Interestingly, this large body of QCVV work cannot usually be applied to the assessment of quantum annealing (QA) computers, such as the quantum devices developed by D-Wave Systems \cite{Johnson2011,dwave_docs}.
The fundamental challenge in conducting characterization, verification, and validation of quantum annealing devices (QAVV) is that available hardware platforms only allow measuring the state of the system in a fixed basis (the so-called computational $z$-basis) and at the completion of a specified annealing protocol.
Consequently, the QA user can only observe a fairly limited projection of the quantum state that occurs during the hardware's computation, which raises a variety of challenges for how to best conduct QAVV.

QAVV efforts began in earnest around 2010 with a number of quantum hardware validation efforts that were successful in demonstrating quantum state evolution in small systems with 8 to 20 qubits  \cite{PhysRevB.81.134510,PhysRevX.4.021041,Boixo2013,PhysRevB.87.020502}.\footnote{Note that several of these works require hardware measurements that are not available to the users of current QA platforms.}
After these initial efforts, QAVV has focused almost exclusively on system-level benchmarks that consider transverse field Ising models with 100s to 1000s of qubits \cite{Job_2018}.
These system-level metrics of QA hardware platforms have generally shown strong qualitative agreement with idealized QA simulations \cite{Boixo2014,PhysRevA.91.042314,Harris162,King2018}. 
However, identifying the root causes for the deviations from idealized QA simulations remains an open research topic.

This work is motivated by the observation that QAVV is lacking in component-level metrics that can be used for characterization, verification, and validation of individual components of large QA hardware platforms.
Taking inspiration from the single-qubit error rate metrics developed in the QCVV literature, this work highlights the usefulness of conducting single-qubit fidelity assessment of individual qubits in a QA platform.
The proposed protocol is able to extract key metrics of individual qubits, such as their effective temperature, noise, and bias, and is executed in parallel for all of the qubits, providing insights into the variability of qubit properties across an entire QA device.
Systematically measuring these fine-grained single-qubit properties can assist in the calibration of idealized QA simulations that seek to emulate specific hardware devices and provides several key metrics of tracking technical improvements on QA hardware platforms over time (e.g., in relation to effective temperature and qubit noise properties).

This work begins by introducing the foundations of quantum annealing for a single qubit in Section \ref{sec:single-qubit} and derives an effective single-qubit model that can be reconstructed from the observations of a particular hardware device.
Leveraging this building block, we then propose a full-chip single-qubit assessment protocol for quantum annealing in Section \ref{sec:full-chip} and illustrate how such a protocol can uncover some surprising trends in system-wide qubit performance.
A brief study on different annealing times highlights how qubit performance can be impacted by annealing procedure in Section \ref{sec:annealing-schedule}. Section \ref{sec:conclusion} concludes the paper with a discussion of the usefulness of the proposed protocol and future work.

\section{Single-Qubit Quantum Annealing}
\label{sec:single-qubit}

The foundation of current quantum annealing platforms is the Ising model Hamiltonian \cite{gallavotti2013statistical},
\begin{equation}
H_{\text{Ising}} = \sum_{ij \in E} J_{ij} \widehat{\sigma}^{z}_i \widehat{\sigma}^{z}_j + \sum_{i \in V}  h_{i} \widehat{\sigma}^{z}_i,
\end{equation}
where $V$ is the set of qubits and $E$ is the set of programmable interactions between qubits.
The elementary unit of this model is a qubit $i \in V$ described by the standard vector of Pauli matrices $\{\widehat{\sigma}^x, \widehat{\sigma}^y, \widehat{\sigma}^z \}$ along the three spatial directions $\{x,y,z\}$.
The outcome of the quantum annealing process is specified by a binary variable $\sigma_i$ that takes a value $+1$ or $-1$ and corresponds to the observation of the spin projection in the computational basis denoted by $z$. 
The final state of each qubit is influenced by user-specified values of local fields $h_{i}$ and two-qubit couplers $J_{ij} \; (i,j \in E)$.
This model is interesting because it can readily encode challenging computational problems arising in the study of magnetic materials, machine learning, and optimization \cite{hopfield1982neural,panjwani1995markov,lokhov2018optimal,Kochenberger2014}.

The quantum annealing protocol strives to find the low-energy assignments to a user-specified $H_{\text{Ising}}$ problem by conducting an analog interpolation process of the following transverse field Ising model Hamiltonian:
\begin{equation}
    H(s) = A(s) \sum_{i \in V} \widehat{\sigma}^{x}_i + B(s) H_{\text{Ising}}.
\end{equation}
The interpolation process starts with $s = 0$ and ends with $s = 1$. The two interpolation functions $A(s)$ and $B(s)$ are designed such that $A(0) \gg B(0)$ and $A(1) \ll B(1)$, that is, starting with a Hamiltonian dominated by $\sum_{i \in V}\widehat{\sigma}^{x}_i$ and slowly transitioning to a Hamiltonian dominated by $H_{\text{Ising}}$.
In an idealized setting and when this transition process is sufficiently slow, the quantum annealing is referred to as adiabatic quantum computation.
The adiabatic theorem states that if the interpolation is sufficiently slow and the quantum system is isolated, proposed QA protocol will always find the ground state (i.e., optimal solution) to the $H_{\text{Ising}}$ problem \cite{quant-ph-0001106,PhysRevE.58.5355}.
However, in existing QA hardware platforms, a wide variety of non-ideal properties can impact the results of a QA computation \cite{PhysRevA.91.062320,Boixo2016,Smirnov_2018}.
In particular, the D-Wave hardware documentation discusses five known sources of deviations from an ideal QA system called {\it integrated control errors} (ICE) \cite{dwave_docs}, which include: background susceptibility; flux noise; DAC quantization; I/O system effects; and variable scale across qubits.

In the spirit of conducting QAVV for the smallest possible component of a QA device, this work considers a variant of QA that is restricted to a single qubit.
Specifically, it considers a system of the form
\begin{equation}
    H(s) = A(s) \; \widehat{\sigma}^{x} + B(s) \; h \widehat{\sigma}^{z}.
\end{equation}
Despite the simplicity of this model, the imperfections of real-world QA platforms make it a useful tool for assessing the performance of individual qubits in practice.

\subsection{An Effective Single-Qubit Model}

The measurement outcomes of a single-qubit quantum annealing experiment take the form of a probability distribution over the two possible observable projections $\sigma \in \{-1,+1\}$.
This probability distribution can be fully characterized by a single parameter $h^{\textrm{eff}}$, coined effective field, in the following manner:
\begin{align}
\mathbb{P}\left(\sigma =  \pm1 \right) = \frac{\exp{\left( h^{\textrm{eff}}\sigma\right)}}{2\cosh{h^{\textrm{eff}}}}. \label{eq:effective_outcome_prob}
\end{align}
The value of $h^{\textrm{eff}}$ depends on the experiment's input parameters and is, in particular, a function of the user input field $h$.
In the case of a classical magnet placed into a persistent external magnetic field $h$ in a thermal equilibrium at temperature $\beta^{-1}$, one will observe a linear relationship between the output and input fields of the form $h^{\textrm{eff}} = \beta h$.
This linear mapping is called a \emph{classical} Gibbs distribution for a single spin. 
However, it was observed in \cite{2012.08827} that the ICE effects of available QA hardware platforms result in an input/output relationship that is more complicated and is better described by a \emph{mixture} of \emph{quantum} Gibbs distributions, which is a generalization of its classical counterpart. 
The derivation provided in \cite{2012.08827} proposes the following mixture of canonical density matrices,
\begin{align}
\rho &= \frac{1}{2}\sum_{s=\pm 1}\frac{\exp\left( \beta ( \gamma h \widehat{\sigma}_x + (h + b + \eta s) \widehat{\sigma}_z ) \right)}{\Tr{\exp\left( \beta (  \gamma  h \widehat{\sigma}_x + (  h + b + \eta s) \widehat{\sigma}_z ) \right)}},
\label{eq:mixture_density_matrix}
\end{align}
which describes a quantum spin in thermal equilibrium at temperature $\beta^{-1}$ subject to a magnetic field with an adjustable component $h$, uncontrollable components for bias $b$, uniform binary noise of magnitude $\eta$, and a transverse field of magnitude $\gamma h$ that is proportional to the input field.
According to the density matrix from Eq.~\eqref{eq:mixture_density_matrix}, the expected value of observing the spin along the $z$-components is given by the standard quantum relation $\mathbb{E}\left(\sigma\right) = \textrm{Tr}\left(\rho \widehat{\sigma} \right)$. Combining this expression with Eq.~\eqref{eq:effective_outcome_prob} results in the input/output field model,

\begin{align}
\tanh & (h^{\text{eff}}) = (h + b + \eta) \frac{\tanh \left( \beta \sqrt{(\gamma h)^2} + (h+ b + \eta)^2 \right)}{2\sqrt{(\gamma h)^2 + (h+ b + \eta)^2}} \nonumber\\
 + & (h + b - \eta) \frac{\tanh \left( \beta \sqrt{(\gamma h)^2 + (h + b - \eta)^2}\right)}{2\sqrt{(\gamma h)^2 + (h+ b - \eta)^2}},
\label{eq:h_quantum_noise}
\end{align}
which depends on four parameters: the inverse temperature $\beta$, the transverse field gain $\gamma$, the uncontrollable field bias $b$ and standard deviation of the noise $\eta$.
Notice that the model in Eq.~\eqref{eq:h_quantum_noise} reduces to the simple classical Gibbs relationship when $\gamma = b = \eta =0$. 

\begin{figure}[!t]
\centering
\includegraphics[width=\linewidth]{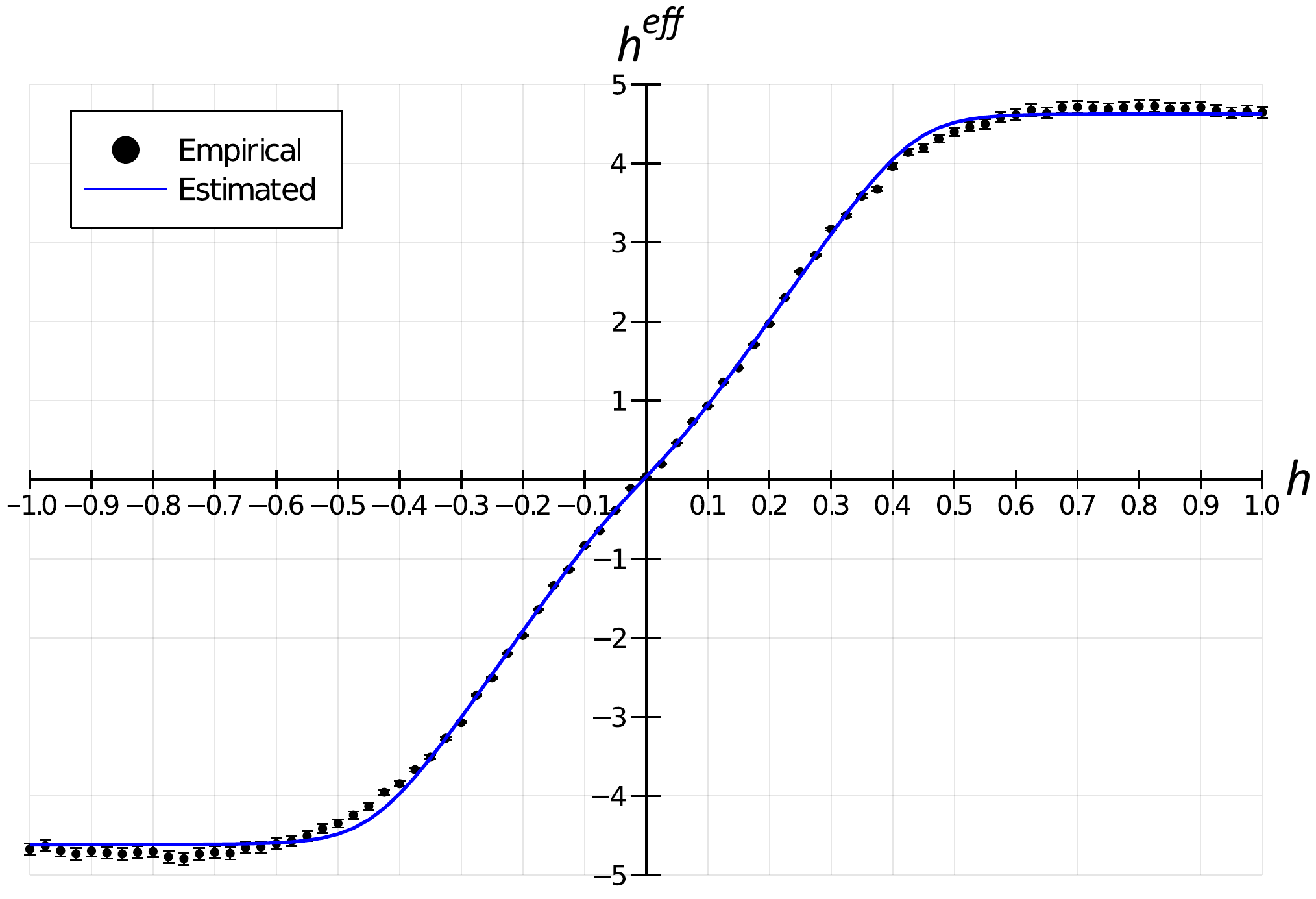}
\caption{Effective output magnetic fields $h^{\text{eff}}$ measured for different input magnetic fields $h$ for a representative qubit ($305$). Each point is estimated using $M = 5 \times 10^6$ samples with a confidence level of $\alpha=0.997$ represented by the error bars. In this case, fitting the effective single-qubit model results in the parameters $\beta = 11.18$, $b = 0.0046$, $\gamma = 0.0196$, and $\eta =  0.0514$, which are reflected by the blue {\it estimated} line.}
\label{fig:sq-example}
\end{figure}

\subsection{Fitting the Single-Qubit Model}

Given that available QA hardware platforms only allow users to observe the system state in the computational basis (i.e., $z$), it is not immediately obvious how one might recover the four model parameters (i.e., $\beta, b, \eta, \gamma$) proposed in Eq.~\eqref{eq:h_quantum_noise}.
A key insight from \cite{2012.08827} is that these four parameters can be inferred from the signatures that appear when running single-qubit annealing for different values of the input field $h$ as described in this section.
At a high level the procedure consists of: (i) selecting a particular set of $h$ values that will be measured on the hardware, denoted by the set $P$; (ii) collecting $M$ samples from the hardware for each $h \in P$, which are used to estimate the empirical mean of the qubit's spin $\sigma$; (iii) using maximum likelihood estimation to recover the best fit values of $\beta, b, \eta, \gamma$ given the observed relationship between $h$ and the empirical mean of the qubit.
The final result of a particular instantiation of this procedure is presented in Figure \ref{fig:sq-example}.

In particular, the model parameters proposed in Eq.~\eqref{eq:h_quantum_noise} can be estimated using the standard Maximum Likelihood Estimation (MLE) approach,
\begin{align}
    \left( \widehat{\beta},\widehat{b},\widehat{\eta},\widehat{\gamma} \right) = \operatorname{argmax}_{\beta,b,\eta,\gamma}  L(\beta,b,\eta,\gamma),
\end{align}
where $L$ is the likelihood function of the four parameters $\beta$, $b$, $\eta$, and $\gamma$.
According to the effective single-qubit model, the probability of observing a configuration $\sigma$ conditioned on a value of the input magnetic field is given by Eq.~\eqref{eq:effective_outcome_prob}, where the effective field $h^{\textrm{eff}}(h,\beta,b,\eta,\gamma)$ depends on the model parameters through Eq.~\eqref{eq:h_quantum_noise}. It is straightforward to derive the following likelihood function for this model,
\begin{align}
    L(\beta,b,\eta,\gamma) = \sum_{h\in P} h^{\textrm{eff}}(h) \widehat{\mathbb{E}}\left[\sigma \mid h\right] - \log \cosh h^{\textrm{eff}}(h),
\end{align}
where $\widehat{\mathbb{E}}\left[\sigma \mid h\right]$ denotes the empirical mean of the spin configuration on the input field $h\in P$.
This likelihood function can easily be maximized using established numerical optimization techniques yielding the best fit parameters for the model.

It is important to briefly remark on the data requirements for an accurate estimation of the model parameters that are encoding subtle variations of $h^{\textrm{eff}}$, especially at large $h$ values.
Note that, for a particular value of $h$, one collects $M$ samples to extract a conditional expectation $\widehat{\mathbb{E}}\left[\sigma \mid h\right]$, which corresponds to an empirical effective field $h^{\textrm{eff}} = \arctanh \widehat{\mathbb{E}}\left[\sigma \mid h\right]$.
For a particular value of $M$, this estimator is subject to an accuracy limit due to finite sampling.
For large values of $\vert h^{\textrm{eff}} \vert$, the probability of observing a qubit misaligned with the effective field decreases exponentially with the field's intensity; see Eq.~\eqref{eq:effective_outcome_prob}.
Therefore, if $h^{\textrm{eff}} = 5$ one only expects to see 1 misaligned spin configuration in every $22,000$ observations, requiring millions of samples to have a confident estimation of $h^{\textrm{eff}}$.
It is hence necessary to adjust these data collection requirements to be consistent with the QA hardware's performance.
This finite sampling accuracy challenge is addressed in this work by setting $M$ to a level that provides tight confidence intervals ($\sigma = 0.997$) around the estimation of $h^{\textrm{eff}}$ for the particular QA hardware that was considered, which resulted in $M = 5 \times 10^6$.

\begin{figure*} [t]
  \centering
  \begin{subfigure}[b]{0.49\textwidth}
      \includegraphics[width=\textwidth]{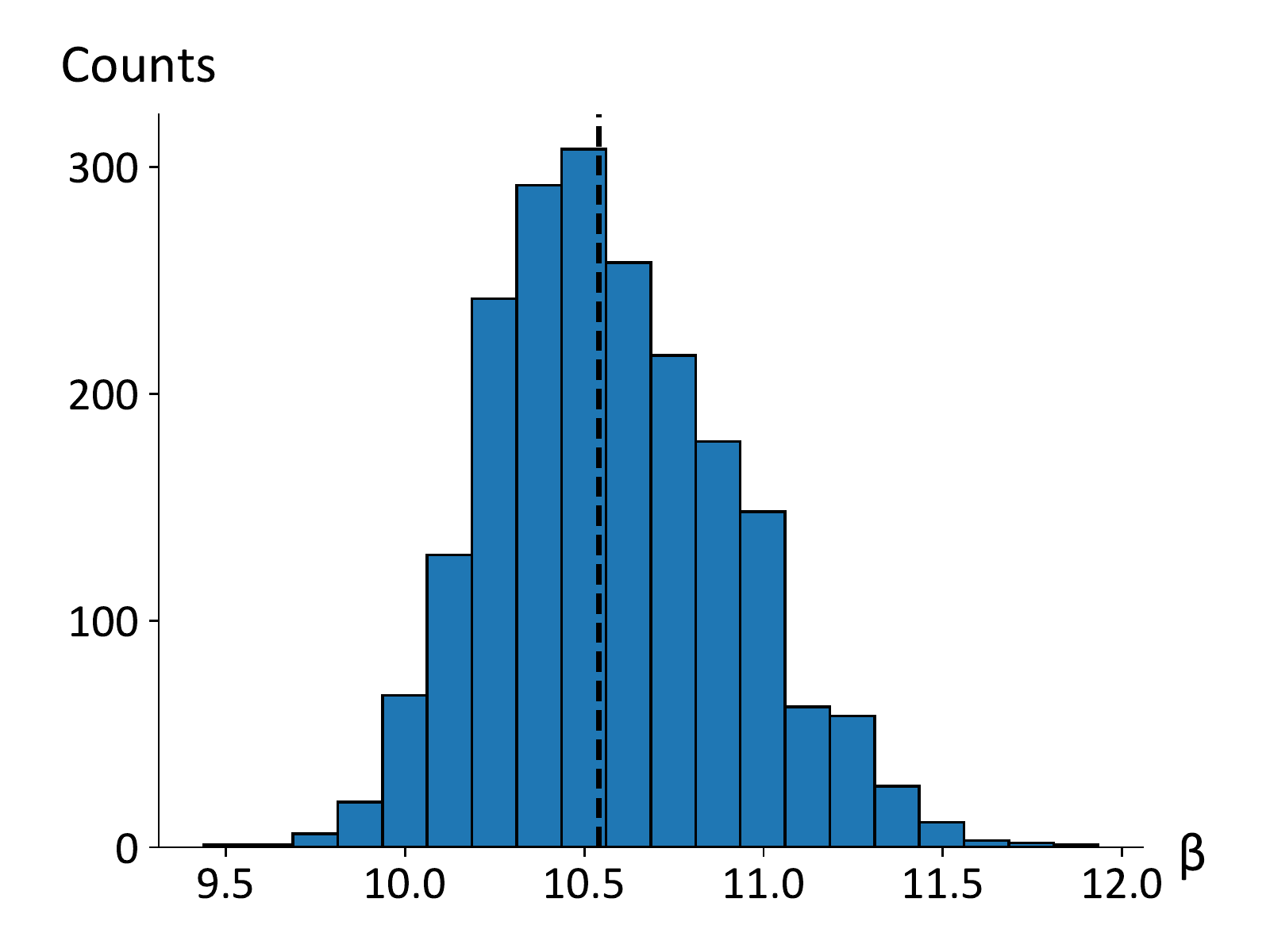}
      \caption{Distribution of $\beta$ over entire chip}
      \label{fig:sq-dist-beta}
  \end{subfigure}
  \begin{subfigure}[b]{0.49\textwidth}
      \includegraphics[width=\textwidth]{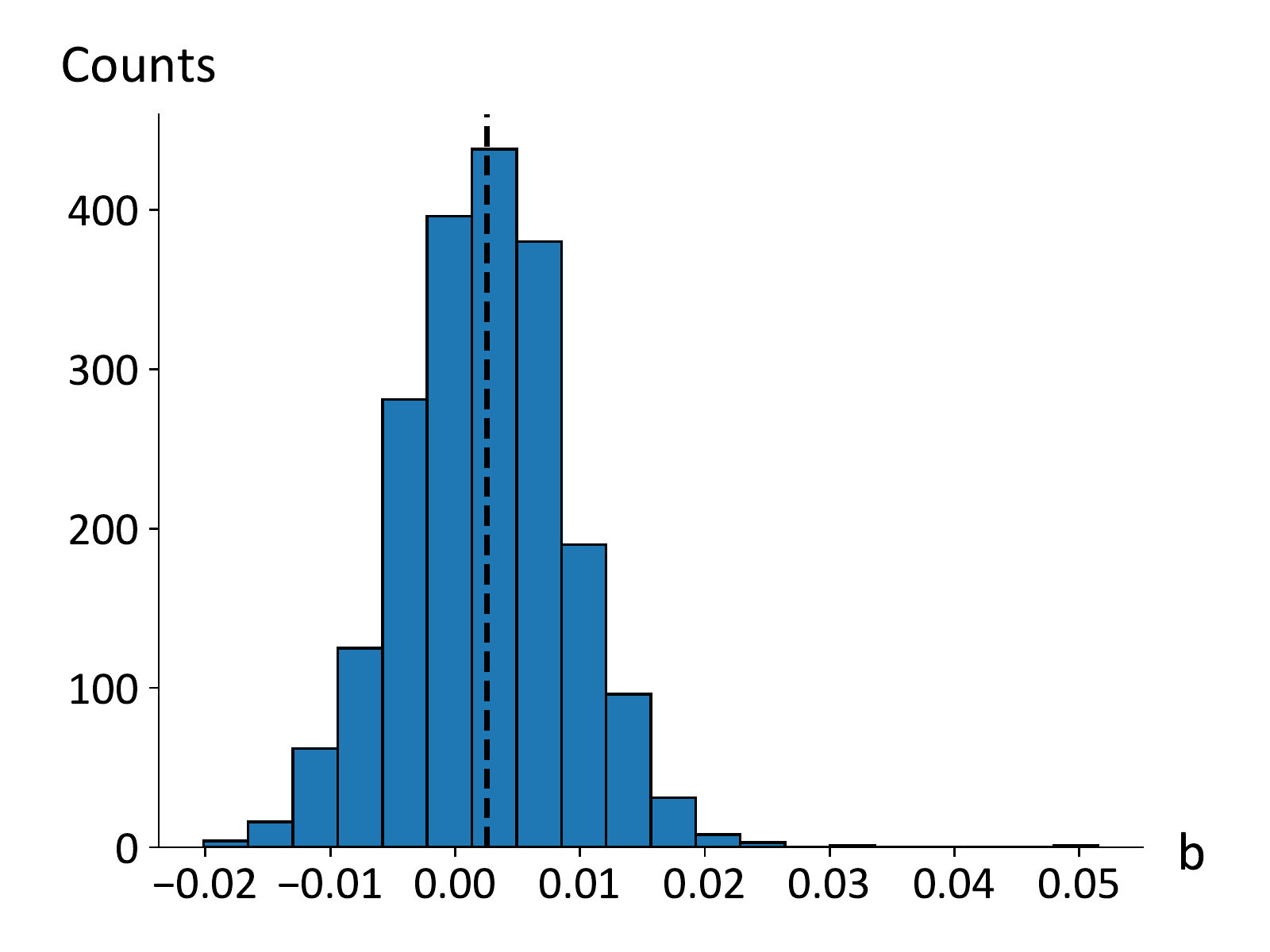}
      \caption{Distribution of $b$ over entire chip}
  \end{subfigure}\\
  \begin{subfigure}[b]{0.49\textwidth}
      \includegraphics[width=\textwidth]{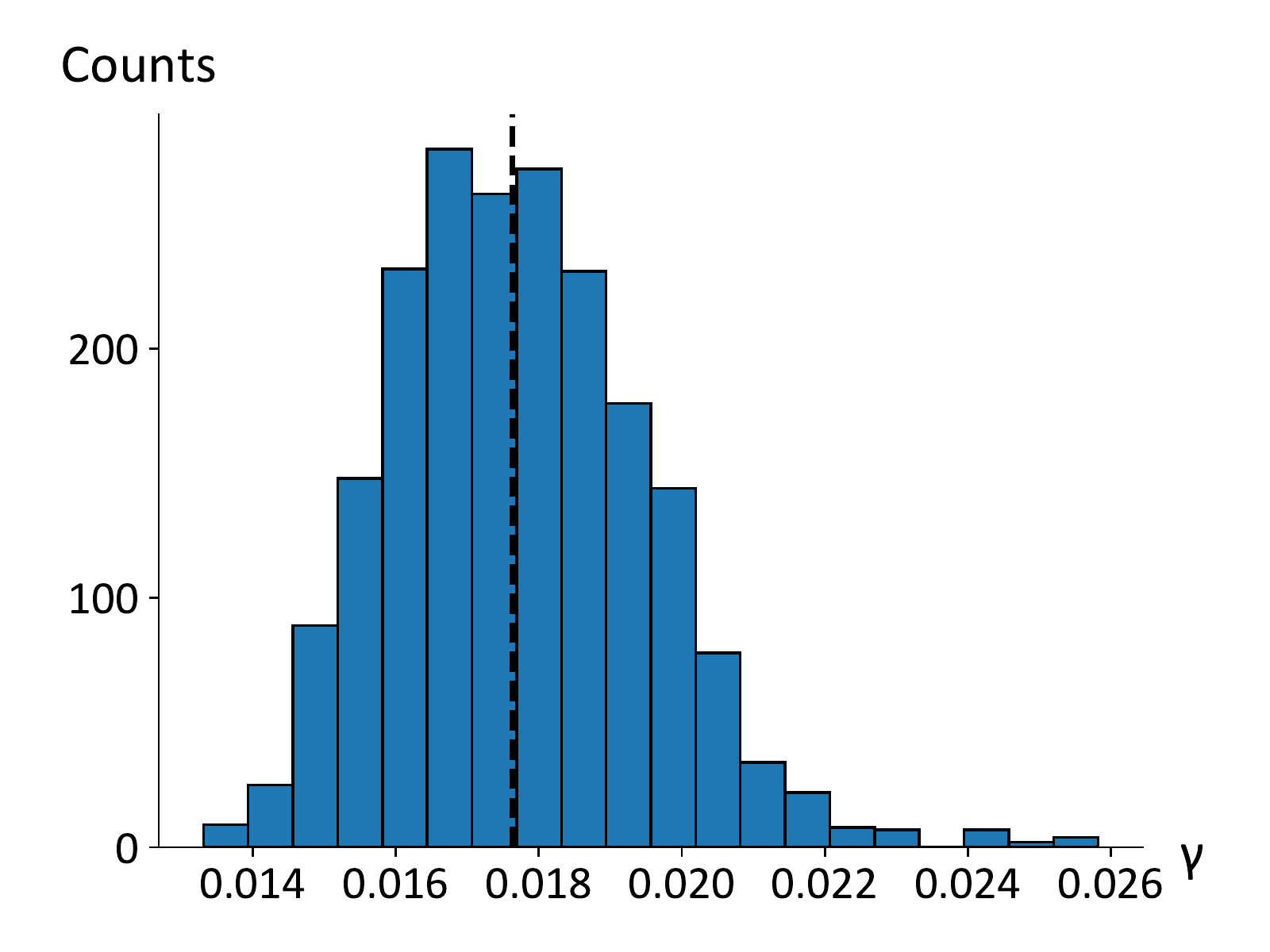}
      \caption{Distribution of $\gamma$ over entire chip}
  \end{subfigure}
   \begin{subfigure}[b]{0.49\textwidth}
      \includegraphics[width=\textwidth]{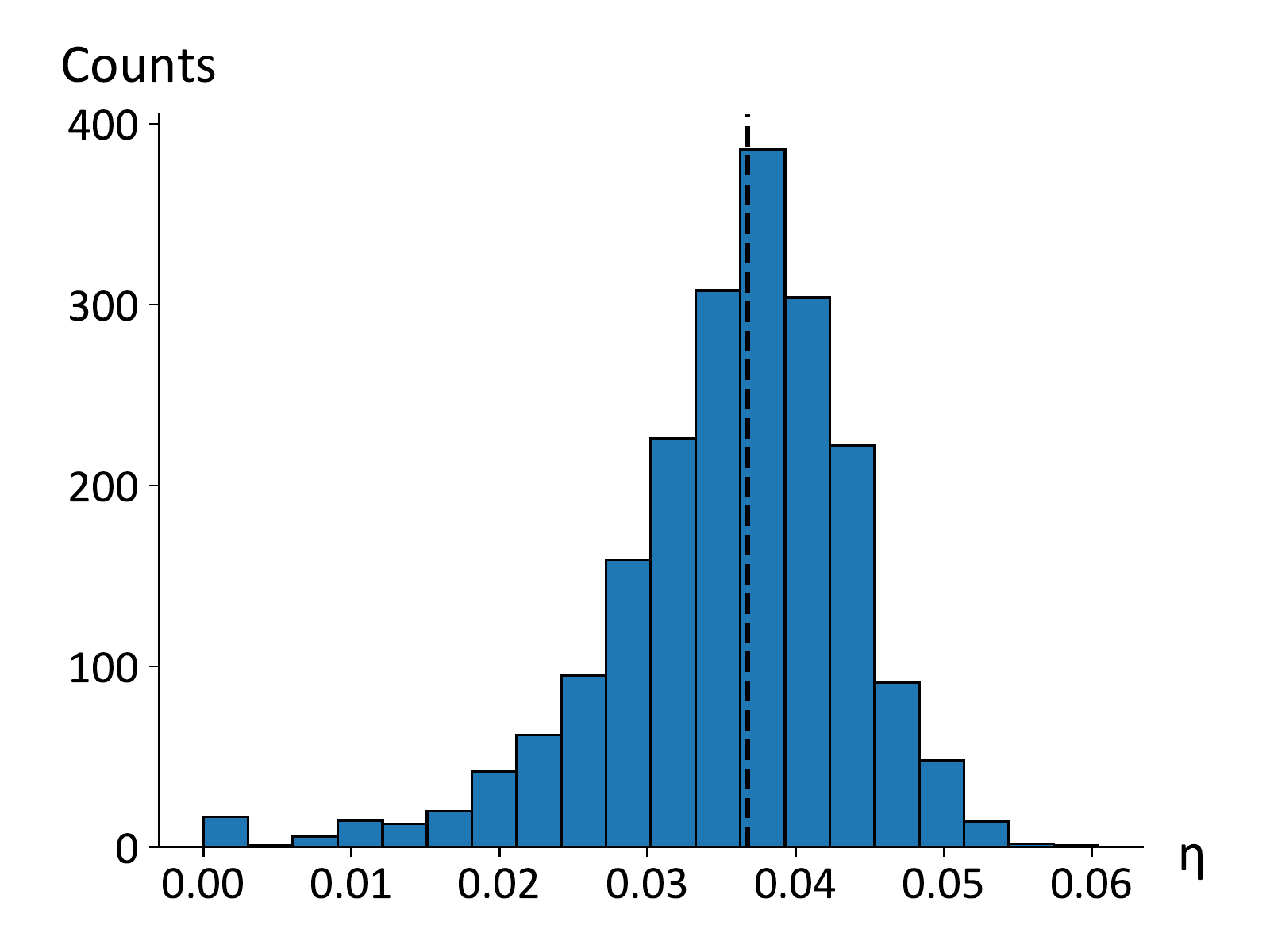}
      \caption{Distribution of $\eta$ over entire chip}
  \end{subfigure}
  \caption{Distributions of the effective single-qubit parameters recovered by the QASA protocol for all 2032 qubits of \texttt{DW\_2000Q\_LANL} system. The dashed line represents median values of each parameter: $\beta = 10.54$, $b = 0.0025$, $\gamma = 0.0176$, and $\eta = 0.0367$. The variability across the qubit population highlights and quantifies the heterogeneity of individual qubits in the hardware.  For QA hardware users, the specific qubit parameters can aid in the calibration and debugging of specific input models. While in the context of QAVV, these distributions can be used to track hardware performance improvements across successive generations of systems.}
  \label{fig:sq-dist}
\end{figure*}

\subsection{Illustration on a Typical Qubit}

To make this single-qubit model-fitting procedure concrete, Figure \ref{fig:sq-example} provides an example of performing the complete procedure on a representative qubit in a 2000Q D-Wave quantum annealing computer.
In this example $81$ input fields (i.e., $h$) ranging from $-1$ to $+1$ are collected and used to recover the effective qubit model parameters using the MLE approach.
The tight error bars on the points indicate that the $h^{\text{eff}}$ values are recovered to a high accuracy, and the close alignment of the best-fit model (blue line) with the observed data illustrates how the effective single-qubit model is able to replicate the key features of the data.
Intuitively, the effective inverse-temperature term $\beta$ sets the general slope of the input/output relationship between $h$ and $h^{\text{eff}}$, the bias term $b$ enables the model to not cross exactly through the origin, the noise term $\eta$ has the effect of lowering the slope near the origin, and the $\gamma$ term flattens the $h^{\text{eff}}$ curve for values of $h$ above $0.4$.
It is important to emphasize that this model (i.e., Eq.~\eqref{eq:h_quantum_noise}) is only an {\it effective} model of the output distribution.
It is not possible to definitively conclude from this experiment the underlying physical cause of these behaviors; however, it is clear that this model, with a transverse field component, is able to statistically reproduce what experimental observations show.

\section{Full-Chip Qubit Fidelity Assessment}
\label{sec:full-chip}

The central insight of this work is that the data collection procedure required for fitting the effective single-qubit model discussed in the previous section can be executed in parallel for every qubit in a QA hardware device.
Consequently, we propose a Quantum Annealing Single-qubit Assessment (QASA) protocol that allows for a detailed characterization of the distribution of effective temperature, offset, noise, and saturation, across an entire QA hardware device.
This enables QA users to quickly verify the level of consistency across the hardware's qubits and to avoid or compensate for non-ideal qubits, adding a new procedure to the QAVV toolbox.

\begin{figure*}[t]
  \centering
  \begin{subfigure}[b]{0.49\textwidth}
      \includegraphics[width=\textwidth]{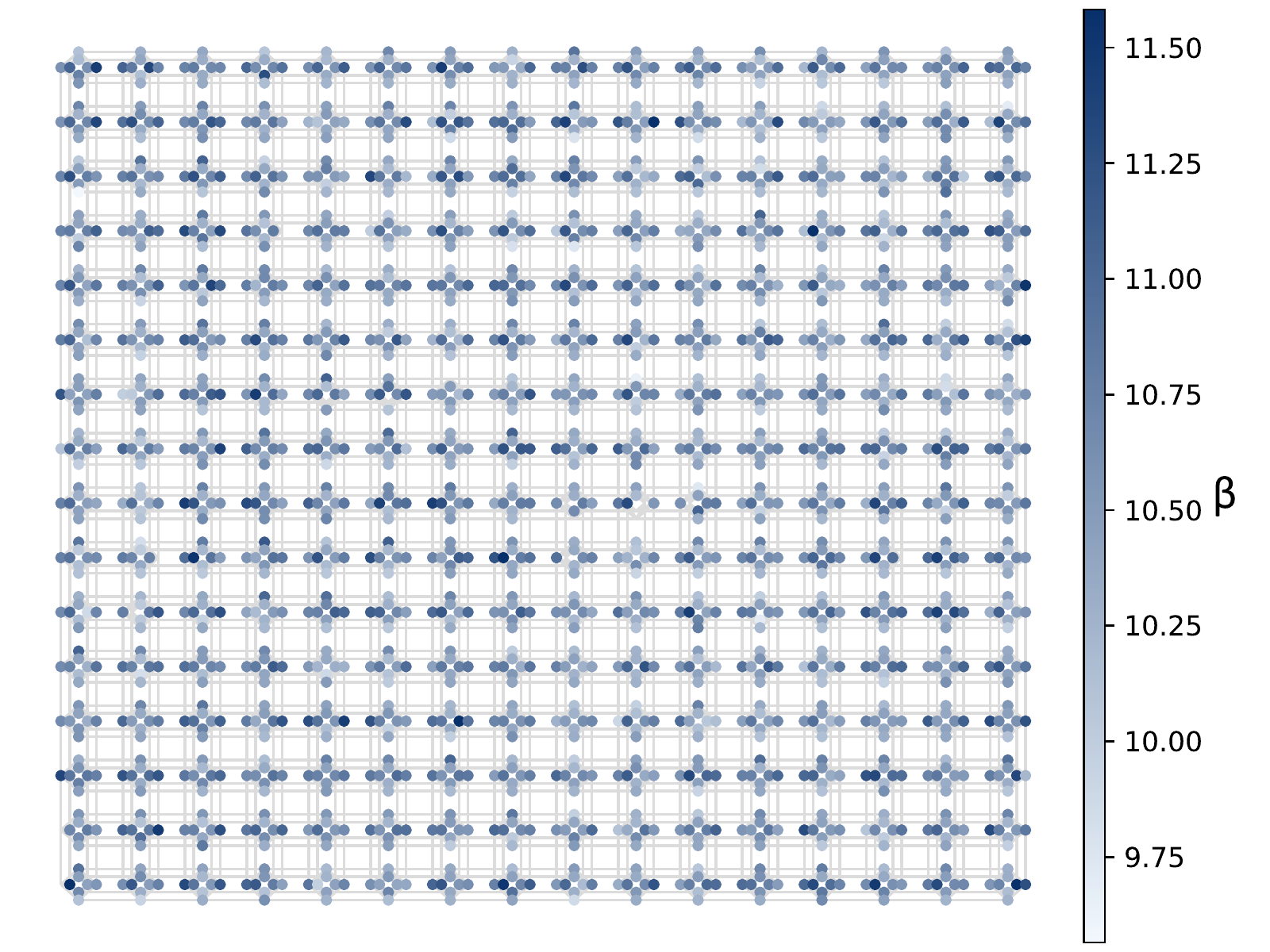}
      \caption{Distribution of $\beta$ over entire chip}
      \label{fig:sq-layout-beta}
  \end{subfigure}
  \begin{subfigure}[b]{0.49\textwidth}
      \includegraphics[width=\textwidth]{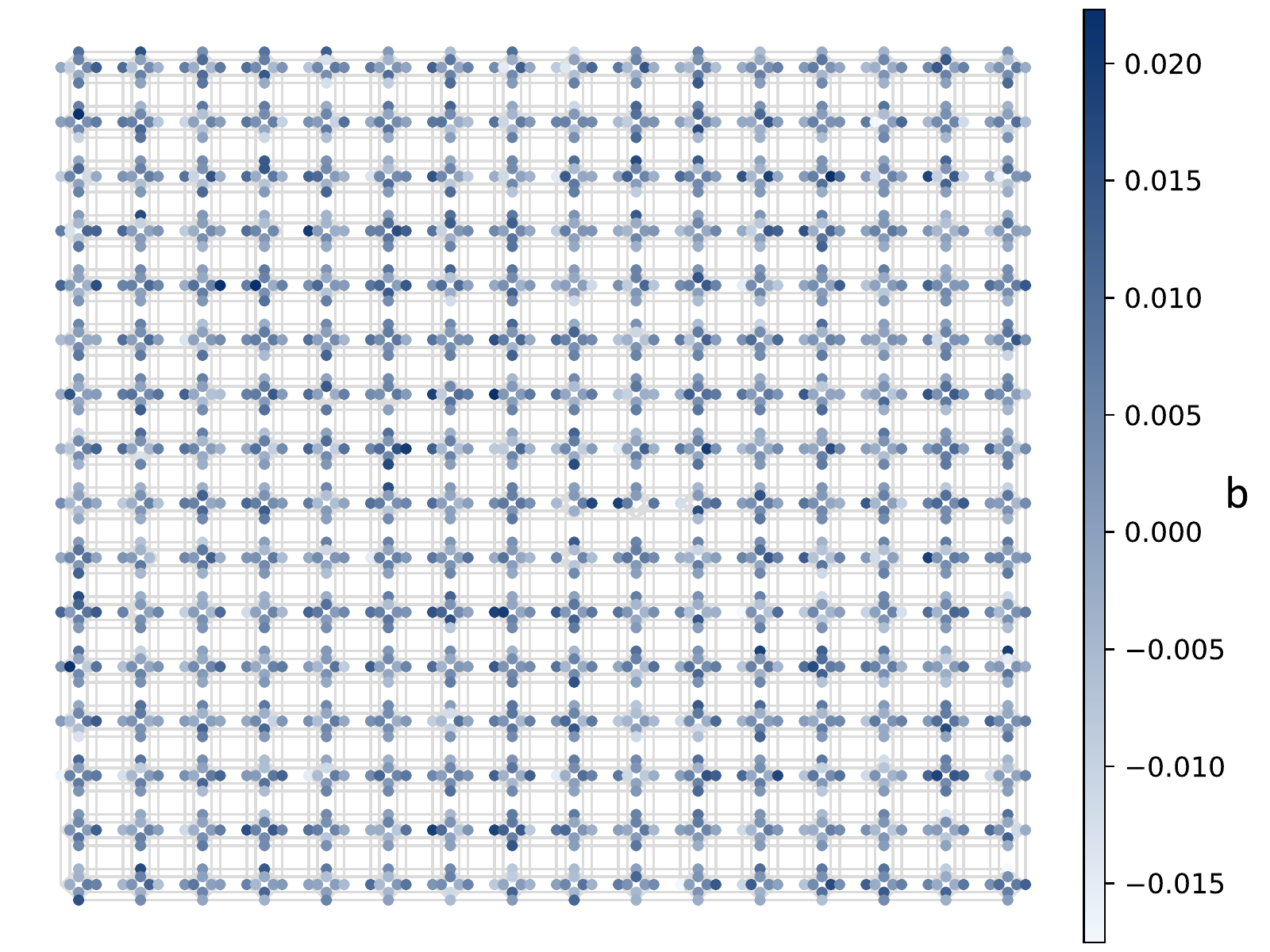}
      \caption{Distribution of $b$ over entire chip}
  \end{subfigure}\\
  \begin{subfigure}[b]{0.49\textwidth}
      \includegraphics[width=\textwidth]{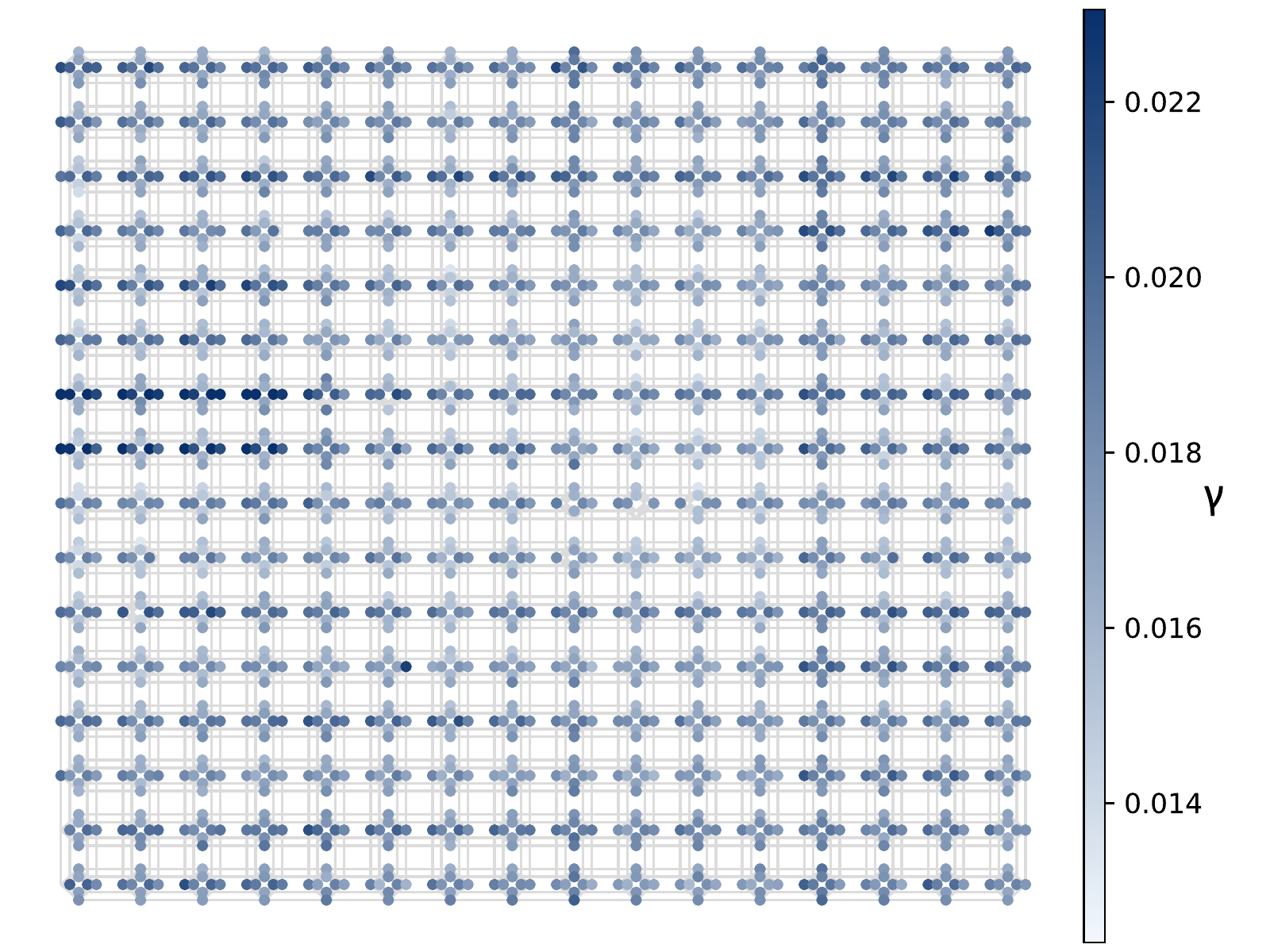}
      \caption{Distribution of $\gamma$ over entire chip}
      \label{fig:sq-layout-gamma}
  \end{subfigure}
   \begin{subfigure}[b]{0.49\textwidth}
      \includegraphics[width=\textwidth]{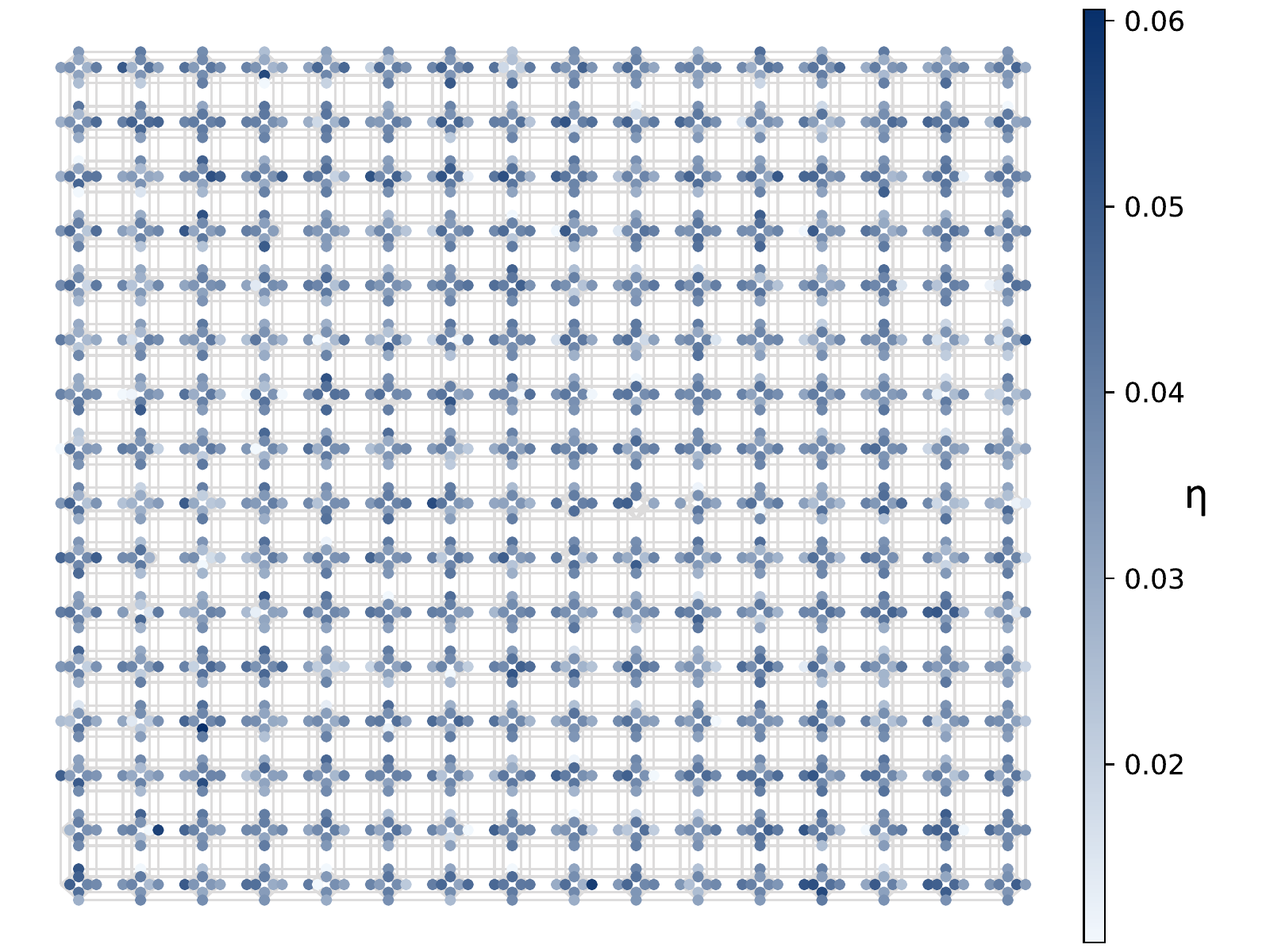}
      \caption{Distribution of $\eta$ over entire chip}
  \end{subfigure}
  \caption{Spacial distributions of the effective single-qubit parameters recovered by the QASA protocol for all 2032 qubits of \texttt{DW\_2000Q\_LANL} system. In this layout, unit cells are depicted as crosses with each horizontal qubit sharing a coupler with each vertical qubit in its cell. Each qubit is colored according to the value of its estimated parameter. Note the horizontal streaks in the $\gamma$ panel, which suggest a higher $\gamma$ value for qubits aligned horizontally rather than vertically. Although it is not as obvious to see, it does appear that the horizontal qubits tend to have higher $\beta$ values as well.
  This is contrasted with the sporadic spacial distribution of $b$ and $\eta$ across the chip, which do not have any particular regional dependence or connection to horizontal and vertical qubits. This heterogeneity between horizontal and vertical qubits highlights a novel QAVV feature that is revealed by the full-chip assessment using the QASA protocol.}
  \label{fig:sq-layout}
\end{figure*}

To demonstrate the efficacy of the QASA protocol for QAVV, this work analyzes a D-Wave 2000Q Quantum Annealer located at Los Alamos National Laboratory, known as \texttt{DW\_2000Q\_LANL}.
This system implements a ${\cal C}_{16}$ chimera graph \cite{6802426}, which consists of a $16 \times 16$ grid of unit cells each containing 8 qubits (4 horizontal and 4 vertical), as illustrated in Figure \ref{fig:sq-layout}.
This architecture supports a maximum of 2048 qubits but this particular system only contains 2032 operational qubits, as the qubit yield in any current D-Wave device is around 99\%.
The system operates at a mean temperature around 15 mK, although this value fluctuates somewhat over time \cite{PhysRevApplied.8.064025}.
Throughout this work the output statistics are collected for 81 $h$ input values in the range of $[-1, 1]$ with a uniform step size of $0.025$.
The following annealing parameters are used unless specified otherwise: {\it flux drift compensation} is disabled, which prevents automatic corrections to input fields based on a calibration procedure that is run a few times each hour; the {\it num reads} is set to 10000, specifying the number of identical executions performed for a single programming cycle of the chip; and the {\it annealing time} is set to 1 $\mu s$.
For each $h$ input value, the $h^{\text{eff}}$ is estimated with $5 \times 10^6$ identical executions to ensure high accuracy of the $h^{\text{eff}}$ estimation.
The following sections discuss the results of running this QAVV protocol on all 2032 qubits in the \texttt{DW\_2000Q\_LANL} system and show that this reveals unique insights into the characterization of this specific device.

\subsection{Effective Qubit Parameter Distributions}

Given that a prevailing assumption of QA modeling is that all of the qubits have identical properties \cite{PhysRevApplied.8.064025,PhysRevA.94.022308,10.3389/fict.2016.00023}, it is important to investigate how reasonable this assumption is in practice.
To that end, we begin by investigating the distribution of parameters output by the QASA protocol.
Figure \ref{fig:sq-dist} presents the empirical distributions of individual qubit parameters across the entire \texttt{DW\_2000Q\_LANL} system.
The first observation of these results is that there is a notable amount of heterogeneity in all of the recovered parameters across the qubits in the hardware.
The second observation is that the variability in the $\beta$ parameter is particularly notable as, in the effective qubit model, $\beta$ is a scaling parameter occurring in the exponent of the density matrix (i.e., $e^{\beta(\dots)}$).
Hence, relatively small changes in $\beta$ can have a dramatic impact on the output statistics.
It is possible that accounting for these variations in the $\beta$ values of different qubits could improve the accuracy of encoding practical problems into the hardware.

When looking closely at the distributions of $\gamma$ and $\eta$ in Figure \ref{fig:sq-dist}, one can observe a slight skew in these parameters relative to a symmetric distribution.
This highlights the potential of QASA protocol to identify outlier qubits, which may be preferable to avoid in applications seeking the best possible consistency or accuracy.
Indeed, a deeper investigation into the $\gamma$ outliers  identifies a particular area of the \texttt{DW\_2000Q\_LANL} system where the system is non-homogeneous; see the darkest values in Figure \ref{fig:sq-layout-gamma}.
One can also notice a few qubits in the $\eta$ distribution that have very low noise values (i.e., $\eta < 0.005$).
It is important to note that the spacing of the $h$ values that are employed in the QASA protocol determines the minimum level of noise that is detectable.
Intuitively, the $h$ values that appear in Figure \ref{fig:sq-example} must be spaced in a way that can detect a slight slope change near the origin to recover a suitable noise value. Very low noise can be mistaken for zero noise, if the slope change is too small to be accurately detected.
In this work we generally found that a $h$ spacing of $0.025$ was sufficient to accurately recover the noise occurring in the \texttt{DW\_2000Q\_LANL} system.
However, as QA hardware continues to improve, the spacing or density of data collection points may need to be adjusted to accurately measure finer noise effects.

\subsection{Spacial Correlation}

Given that there is some variability in the qubit parameters across the hardware platform, a natural follow-on investigation is whether these fluctuations have any dependence on the positioning in the system. 
To that end, Figure \ref{fig:sq-layout} presents the QASA results as a heat-map on a hardware layout of the chip where the qubit color indicates the value of each qubit's parameter.
It is important to highlight that this diagram of the hardware's implementation is a dramatic simplification of the physical implementation where, for example, each qubit (represented by a node in this illustration) is implemented as a superconducting loop connected to a wide variety of control circuitry \cite{Johnson2011,6802426}.
The first observation one can make from this spacial analysis is that there is no immediately obvious correlation in the recovered parameters of $b$ and $\eta$.
In particular, it does not appear that the chip is partitioned into cooler and warmer areas, which would be indicated by a spacial correlation in the $\beta$ parameter.

The most intriguing property revealed by Figure \ref{fig:sq-layout} is the appearance of horizontal and vertical stripes in the $\gamma$ parameter.
This particular structure was not anticipated by any previous work that we are aware of and is a novel insight made possible by the QASA protocol.
Although pinpointing the root cause of these distinctions is outside the scope of this work, it seems likely that this effect is an artifact from some aspect of the hardware's implementation, which is not readily available to QA users.

\subsection{Horizontal and Vertical Qubit Groups}

\begin{figure*}[t]
  \centering
  \begin{subfigure}[b]{0.49\textwidth}
      \includegraphics[width=\textwidth]{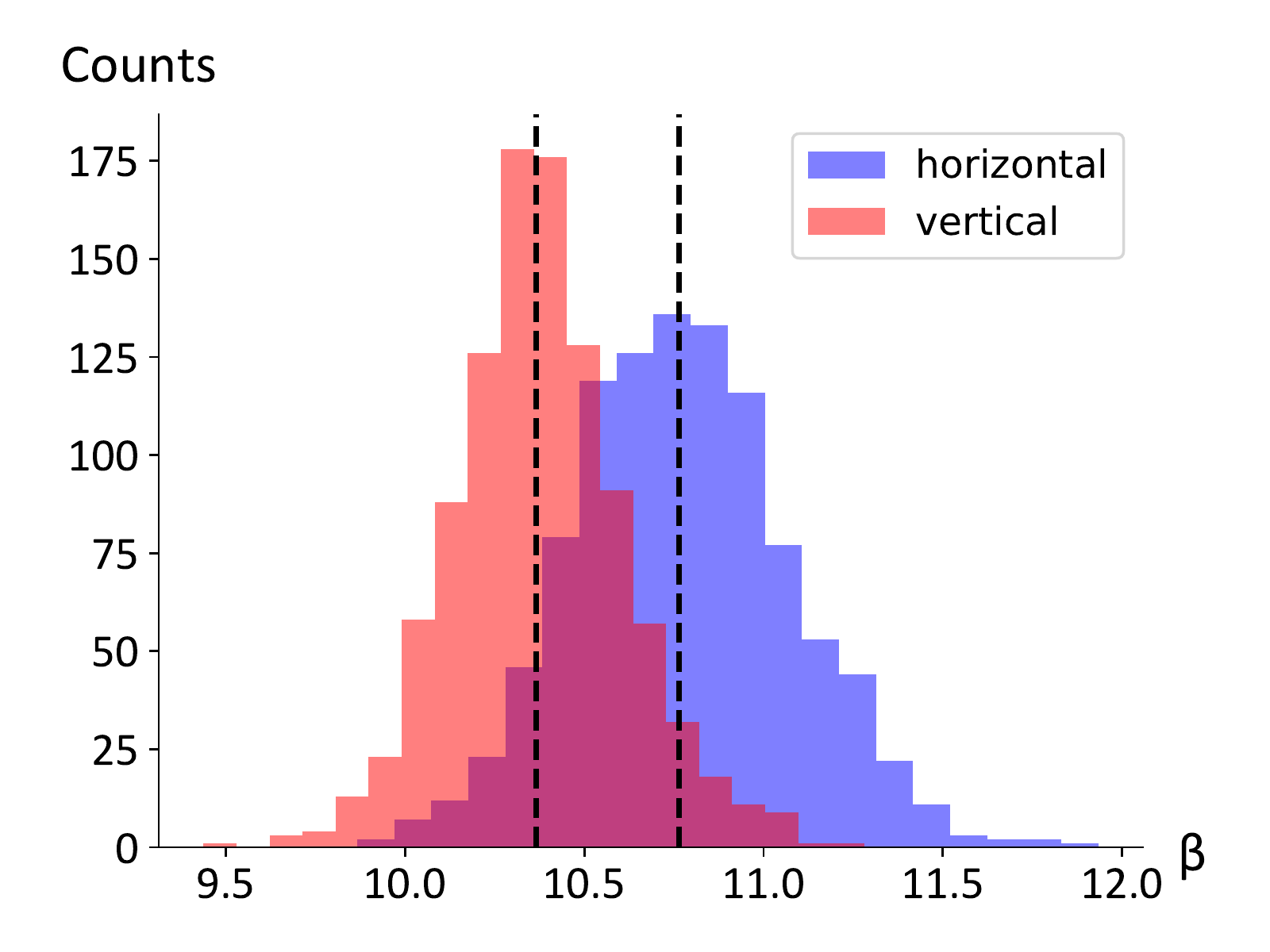}
      \caption{Histograms with respect to $\beta$ over entire chip}
      \label{fig:sq-dist-hv-beta}
  \end{subfigure}
  \begin{subfigure}[b]{0.49\textwidth}
      \includegraphics[width=\textwidth]{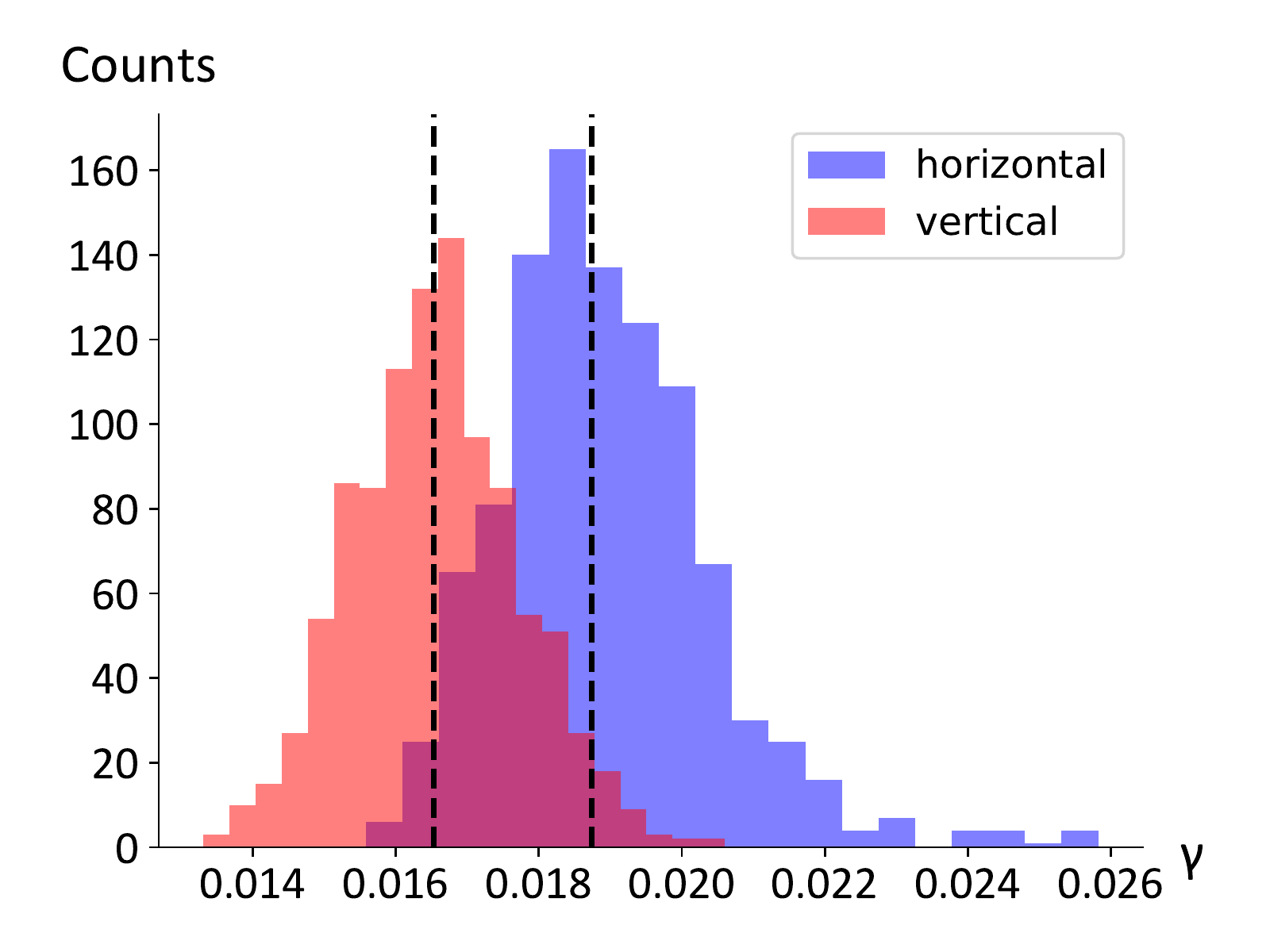}
      \caption{Histograms with respect to $\gamma$ over entire chip}
      \label{fig:sq-dist-hv-gamma}
  \end{subfigure}
  \caption{Distributions of the $\gamma$ and $\beta$ parameters segmented by whether the qubit is aligned horizontally or vertically in the hardware implementation. This particular segmentation reveals a shift in the medians of each distribution of $\beta^{\text{h}} = 10.76$, $\beta^{\text{v}} = 10.37$ and $\gamma^{\text{h}} = 0.0187$, $\gamma^{\text{v}} = 0.0165$, further verifying and quantifying this unanticipated source of heterogeneity in the \texttt{DW\_2000Q\_LANL} system.}
  \label{fig:sq-dist-hv}
\end{figure*}

Inspired by the horizontal and vertical banding appearing in Figure \ref{fig:sq-layout-gamma}, one's understanding of the qubit parameter distributions can be enhanced by first categorizing the data into two groups, one for the horizontal qubits and another for the vertical qubits in the hardware graph.
Figure \ref{fig:sq-dist-hv} presents these two distributions for the $\beta$ and $\gamma$ parameters.
While both horizontal and vertical distributions for the $\beta$ and $\gamma$ parameters appear to have similar variance, the mean of each distribution is notably higher for the horizontal qubits.
This result further emphasizes the observation from visual inspection of Figure \ref{fig:sq-layout-gamma} that the horizontal qubits have consistently higher $\gamma$ and $\beta$ values in comparison with the vertical qubits.

The notable change in the means of the distributions presented in Figure \ref{fig:sq-dist-hv} suggests a systematic difference in the way that the horizontal and vertical qubits respond to inputs in the physical hardware.
One possible explanation could be related to an asymmetry in the chip's hardware layout or to the details of how global annealing control signals are delivered to the qubits, which are known to be shared among vertical and horizontal qubits \cite{6802426}.
Although we can only speculate on the possible root cause of this phenomenon, the QASA protocol nonetheless provides valuable insight into the heterogeneous features of the hardware platform, identifying key areas for further investigation.

\begin{figure*}[t]
  \centering
  \begin{subfigure}[b]{0.49\textwidth}
      \includegraphics[width=\textwidth]{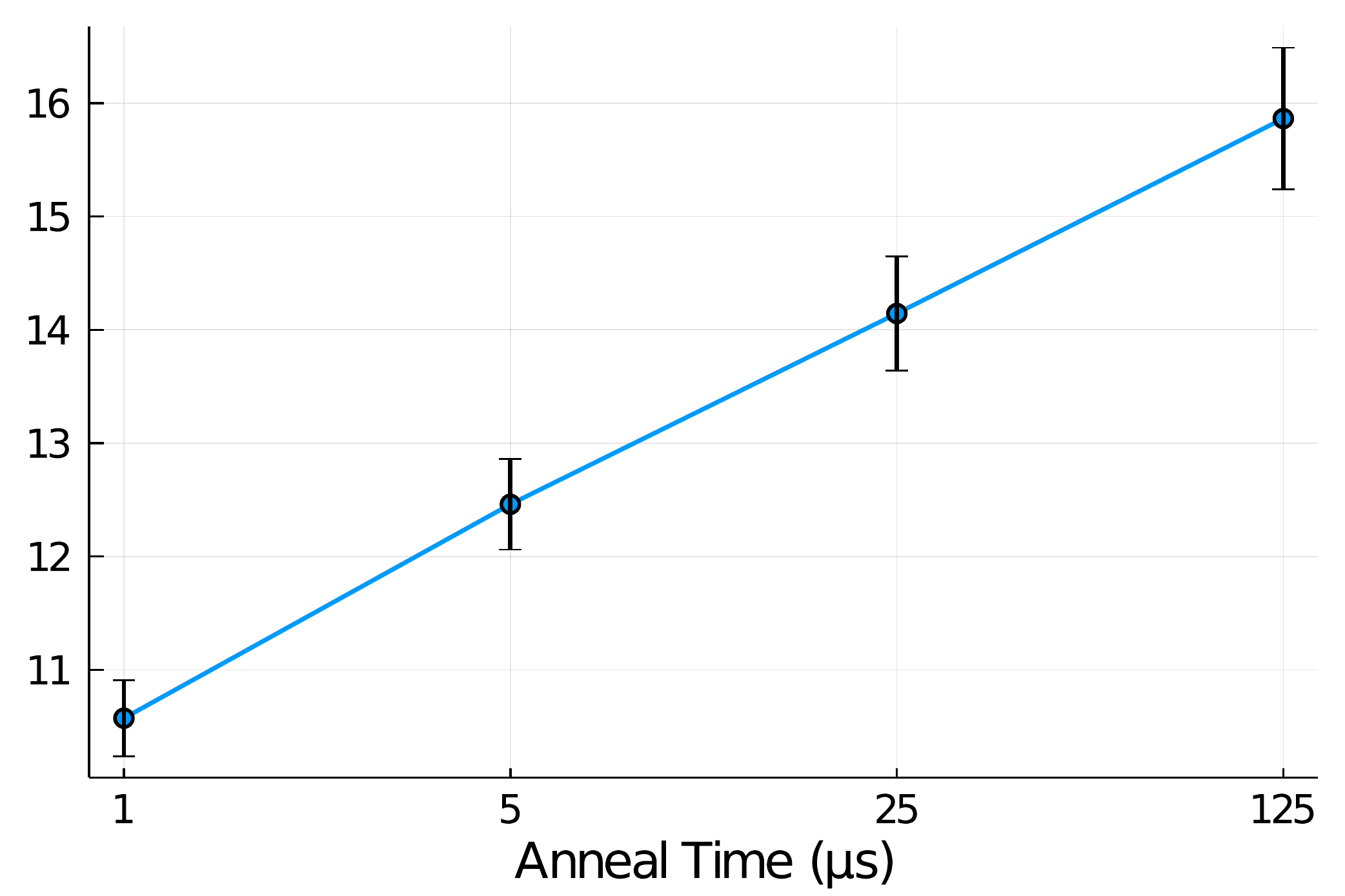}
      \caption{Distribution of $\beta$ dependence on annealing time}
      \label{fig:sq-dist-beta-at}
  \end{subfigure}
  \begin{subfigure}[b]{0.49\textwidth}
      \includegraphics[width=\textwidth]{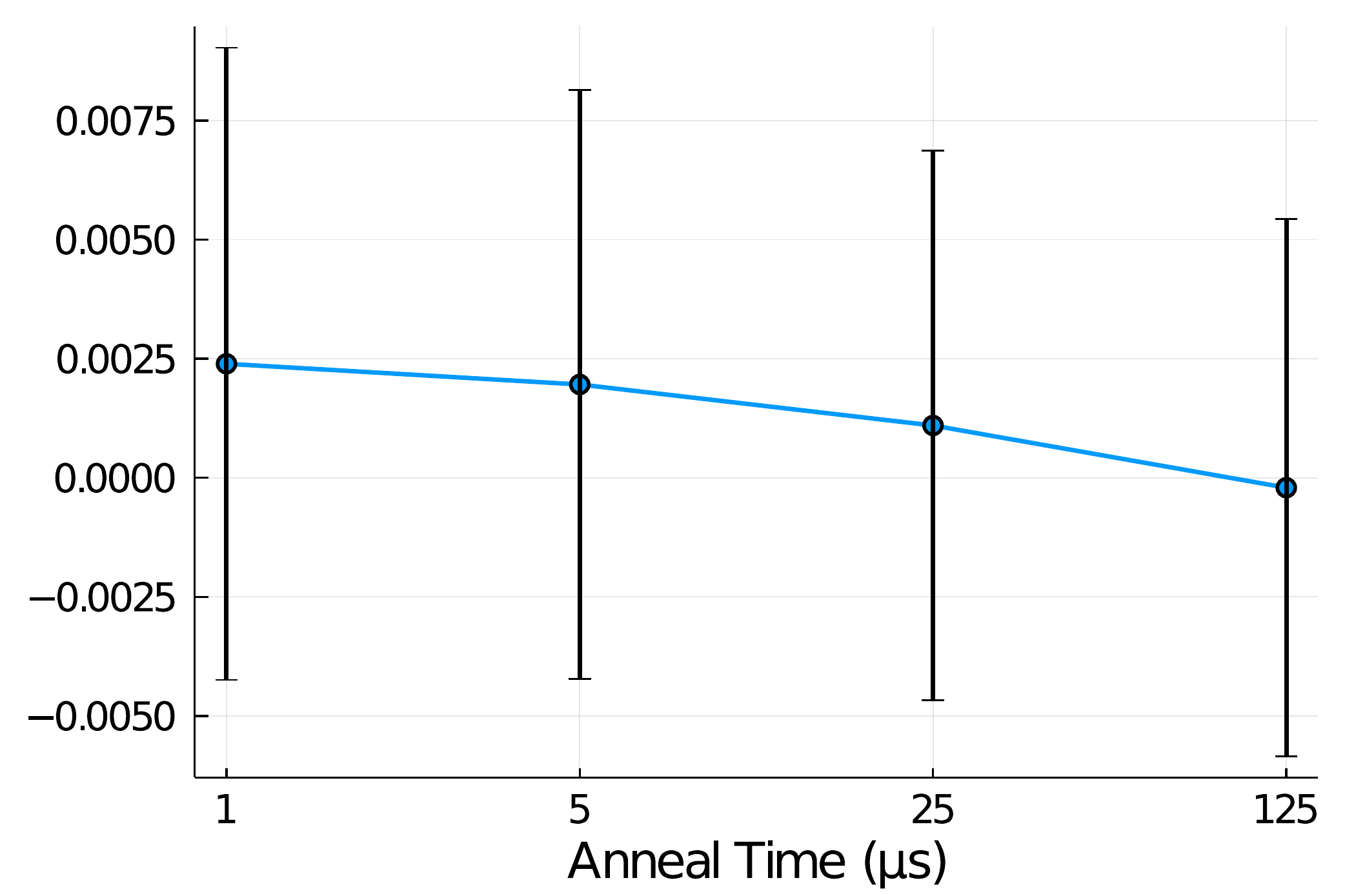}
      \caption{Distribution of $b$ dependence on annealing time}
  \end{subfigure}\\
  \begin{subfigure}[b]{0.49\textwidth}
      \includegraphics[width=\textwidth]{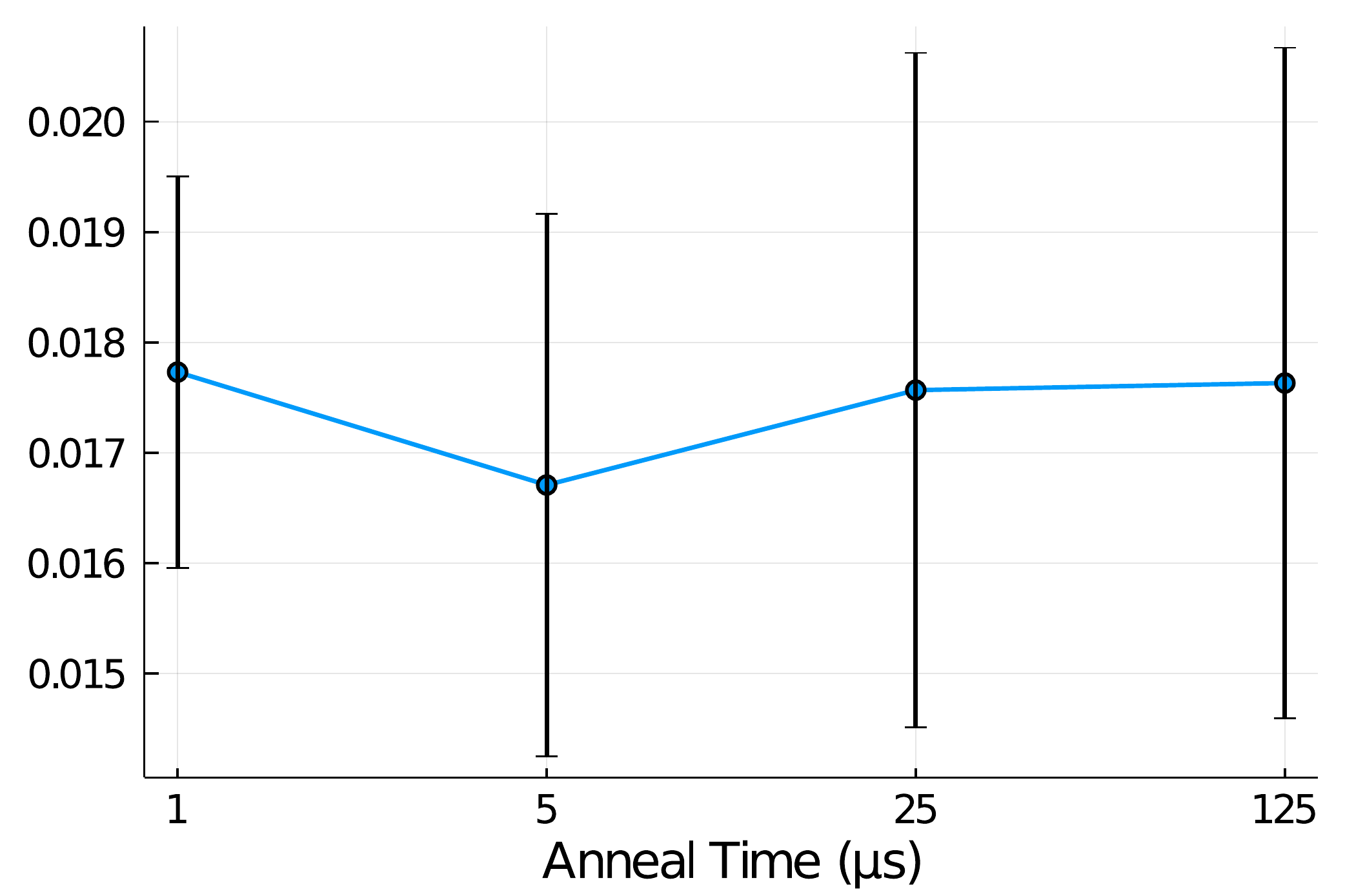}
      \caption{Distribution of $\gamma$ dependence on annealing time}
  \end{subfigure}
   \begin{subfigure}[b]{0.49\textwidth}
      \includegraphics[width=\textwidth]{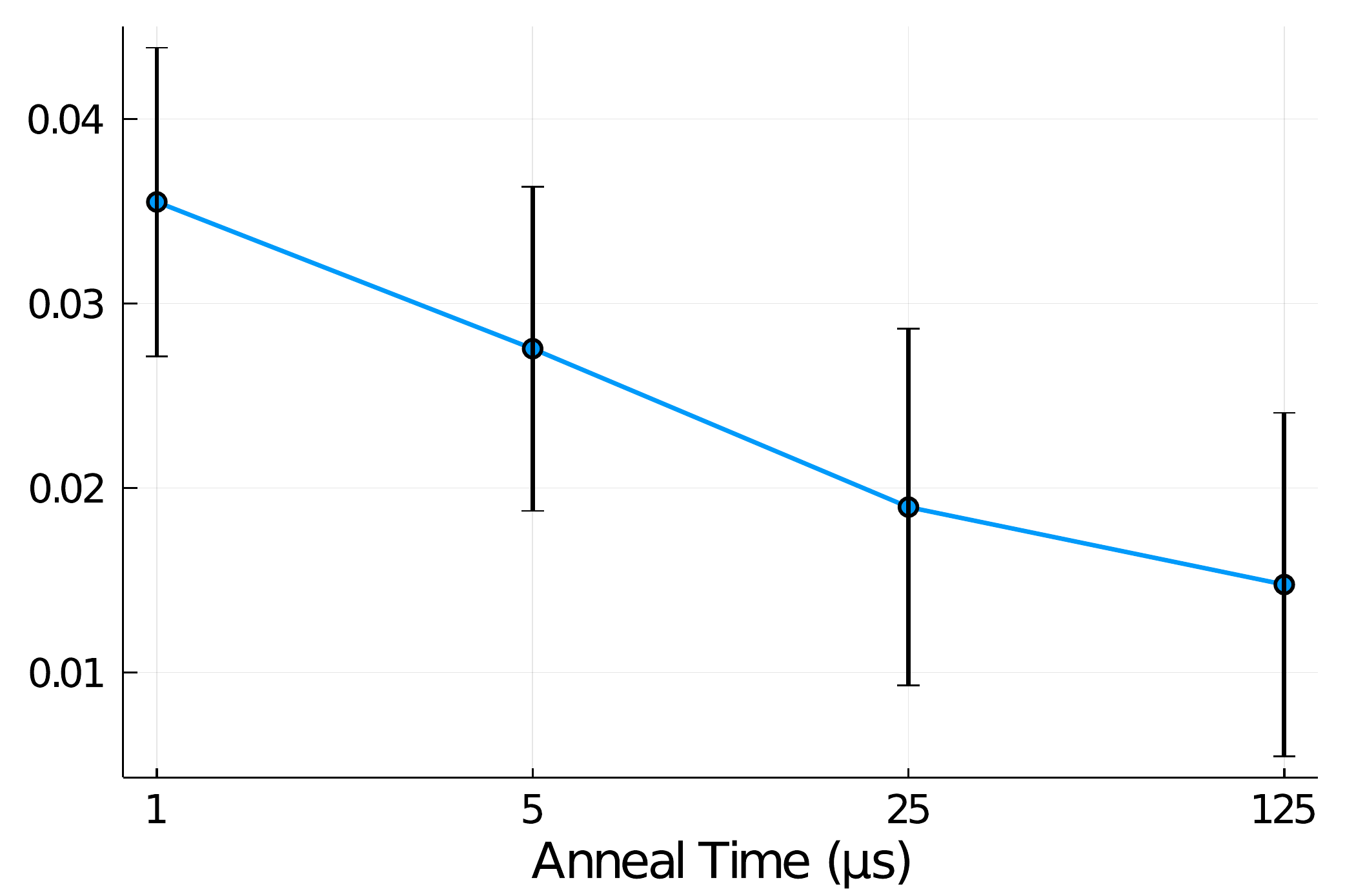}
      \caption{Distribution of $\eta$ dependence on annealing time}
      \label{fig:sq-dist-at-eta}
  \end{subfigure}
  \caption{Impacts of annealing time modulation on single-qubit parameter distributions. For each annealing time, the mean estimated parameter over all 2032 qubits is plotted with whiskers indicating standard deviation of the parameter distribution. $\beta$ has a clear logarithmic dependence on anneal time whereas $b$ and $\gamma$ are well within the standard deviation, showing no relationship to anneal time. There appears to be a slight change in $\eta$ with annealing time, but this is suspected to be an artifact of the specific data collection settings used in this work.} 
  \label{fig:sq-dist-at}
\end{figure*}

\section{Annealing Schedule Impacts}
\label{sec:annealing-schedule}

At first glance the QASA protocol recovers seemingly fundamental properties of the qubits in a QA system (e.g., effective temperature, calibration offsets, $h$-noise). 
It is thus tempting to suggest that these properties are intrinsic to the hardware's implementation.
However, it is important to emphasize that the single-qubit model leveraged by QASA is an effective model describing the statistical properties of a qubit’s behavior.
Real world QA systems are extremely complex devices comprising hundreds of thousands of superconducting electronic components, each of which are influenced by various underlying mechanisms including the annealing scheduling, readout errors, freeze-out, decoherence, excitation, and tunneling \cite{PhysRevA.92.052323,PhysRevA.91.062320,PhysRevApplied.8.064025}.
The goal of QASA is not to characterize the precise details of the quantum dynamical interactions but instead to measure parameters that form an effective model encapsulating the combination of these effects.
To highlight this point, in this section we revisit the QASA protocol while modifying the annealing schedule to show how it can impact the parameters of the effective qubit model.

As an illustrative demonstration, Figure \ref{fig:sq-dist-at} highlights how the QASA parameter distributions can be impacted by different annealing schedules.
The approach is to repeat the QASA protocol while increasing the annealing time by two orders of magnitude, ranging from 1 $\mu s$  to 125 $\mu s$.
It is clear from these results that the parameters for $b$ and $\gamma$ are largely invariant to this particular annealing parameter; however, there is a striking relationship in how $\beta$ increases logarithmically with anneal time, starting at a mean value of $10.5$ and ending with a value of $15.7$.
Intuitively, the dependence of $\beta$ on the annealing time parameter makes sense, as the adiabatic theorem indicates that annealing more slowly increases the likelihood that the system will stay in a ground state of the specified input model. 
This increased preference for ground states is equivalent to a lower effective qubit temperature and therefore a larger $\beta$ value. 
In particular, this result highlights that the effective temperature parameter recovered by the QASA protocol is not held constant by the operating temperature of the hardware device but is instead a feature arising from the complete annealing protocol.

There has been a vigorous debate in the literature around how the observed {\it effective temperature} in a QA device is related to its physical operating temperature \cite{PhysRevApplied.8.064025,10.3389/fict.2016.00023,2012.08827}.
The result presented in this section highlights some of the challenges in using the observed effective temperature for insights into the hardware's operating temperature.
However it is worth noting that the $\beta$ values recovered by the QASA protocol at 1 $\mu s$ are remarkably consistent with the system's measured running temperature of 15 mK \cite{PhysRevApplied.8.064025}.

In particular, it may be reasonable to extrapolate the line in Figure \ref{fig:sq-dist-beta-at} to a value of 0 $\mu s$ to recover a temperature measurement, omitting the impacts of annealing time.
In any case, it is clear that the QASA protocol can provide valuable insight into how the system's physical temperature is connected to an observed effective temperature. 

It is important to briefly mention the $\eta$ parameter's dependence on the anneal time.
A first glance at Figure \ref{fig:sq-dist-at-eta} appears to suggest that $\eta$ decreases with an increased annealing time.
This would be an unexpected outcome from changing the annealing protocol.
After reviewing the results in increased detail, we observed that this trend is due in part to an artifact of the particular points used by the QASA protocol in this work, which has a minimum input resolution of 0.025.
Notice that, as annealing time increases, so does $\beta$, causing the slope of the $h^{\text{eff}}$ curve to increase (see Figure \ref{fig:sq-example}).
This leaves fewer data points in the linear region of the curve, making the detection of subtle slope changes near the origin more challenging, especially for qubits with naturally low noise. 
Although identifying the root cause of this $\eta$ parameter trend requires further investigation, this highlights the possible need to tweak the QASA data collection parameters as QA hardware improves to ensure that sufficient data is collected to recover the key parameters of interest.

Although this section focused on a simple proof-of-concept demonstration using the annealing time parameter,
the other scheduling features of QA hardware such as pausing \cite{PhysRevApplied.14.014100,PhysRevApplied.11.044083}, annealing offsets \cite{PhysRevA.96.042322,Adame_2020}, and custom annealing schedules \cite{King2021,Venturelli2019,10.1371/journal.pone.0244026} all suggest promising avenues for manipulating the effective qubit parameters recovered by the QASA protocol.
To that end, we hope that the QASA protocol can provide a relatively  fast QAVV assessment of how these operational parameters can impact qubit performance in practice.

\section{Conclusion}
\label{sec:conclusion}

Inspired by the effective single-qubit model proposed in \cite{2012.08827}, this work proposed the QASA protocol as a novel tool for conducting QAVV on emerging quantum annealing platforms.
The results derived from running QASA on the \texttt{DW\_2000Q\_LANL} system revealed a number of inconsistencies in qubit performance that were previously unknown, highlighting the usefulness of the proposed approach.
The QASA protocol has further demonstrated its efficacy in this work by revealing, for the first time, an asymmetry in the performance of the qubits from the vertical and horizontal sections of the hardware graph considered herein.
This observation provides a clear point for improving the fidelity and consistency of this particular QA hardware platform.
In time, we hope that the QASA protocol will find a wide range of uses including: tracking the performance improvements of QA hardware platforms, helping hardware designers identify inconsistencies in specific QA devices, and supporting QA users in calibrating algorithms to specific hardware devices.
To support that goal we have released the software that we developed to execute the QASA protocol as open-source, to benefit the broader community in conducting QAVV.

The natural next step for the QASA protocol is to explore how the data collection procedure can be optimized to reduce the amount of chip time required to accurately fit the effective single-qubit model.
In this work we choose a uniform spacing in $h$ with a consistent number of samples for every input value.
Upon conclusion of this work it is now clear that the MLE fitting model would likely benefit from a non-uniform spacing of $h$ that focuses data collections in the areas capturing the most pronounced signatures of $b$, $\eta$ and $\gamma$.
Reducing the number of samples collected at small $h$ values represents another obvious opportunity for reducing the amount of required data collection.

\appendix

\section{}

\subsection{Data Collection and Parameter Recovery Software}

In the interest of making the QASA protocol as widely accessible as possible, the core software for data collection and model parameter fitting is released as open-source software at \url{https://github.com/lanl-ansi/QASA}.
The software consists of two tools: (1) a Python script for extracting data from D-Wave quantum annealing platforms using the Ocean micro-client to collect and combine large numbers of hardware executions; and (2) a Julia-based tool for solving the MLE model for each qubit and building a table of the recovered single-qubit model parameters.
The software is released under a flexible BSD license, which allows for modification, adaptation, and commercial reuse.

\subsection{Raw Data}

The raw data output by QA hardware devices is expensive to acquire and specific to each particular device implementation.
Provided with the supplementary materials of this work are the raw data collected from the \texttt{DW\_2000Q\_LANL} system in the spring of 2021 and the resulting single-qubit model parameters that were recovered from that data.
The data is provided as plain-text in the comma-separated value format (CSV) with a header indicating the value of each column.
In the hardware output data the columns are indicated by: \texttt{h} the input parameter; \texttt{samples} the number of repeated executions for the given input parameter; and \texttt{spin\_id} provides a count of the number of times qubit number \texttt{id} takes the value $-1$ in the output.
In principle, this raw data file can be combined with the open-source software to produce all of the model parameters presented in this work.
However, for convenience, the single-qubit parameters recovered by the MLE model are also provided in the CSV format for analysis without the need of running the software on the provided raw data.

\section*{Acknowledgment}

The authors would like to thank Tameem Albash, Mohammad Amin, Andrew Berkley, and Trevor Lanting for their input on preliminary versions of this work.
The research presented in this work was supported by the Laboratory Directed Research and Development program of Los Alamos National Laboratory under project number 20210114ER and the Center for NonLinear Studies (CNLS).
The computing resources used in this work were provided by the Los Alamos National Laboratory Institutional Computing Program, which is supported by the U.S. Department of Energy National Nuclear Security Administration under Contract No. 89233218CNA000001.

\bibliographystyle{IEEEtran}
\bibliography{main}

LA-UR-21-23013

\end{document}